\documentclass[twocolumn,trackchanges]{aastex7}
\defcitealias{2024Zhong}{\textsc{Paper~I}}
\defcitealias{2025Zhang}{\textsc{Paper~II}}
\defcitealias{2025Zhong_3}{\textsc{Paper~III}}
\usepackage{amsmath}
\begin{document}

\title{Irradiated Atmospheres IV: Effect of Mixing Heat Flux on Chemistry}

\correspondingauthor{Cong Yu}
\email{yucong@mail.sysu.edu.cn}

\author[0009-0004-1986-2185]{Zhen-Tai Zhang}
\affiliation{School of Physics and Astronomy, Sun Yat-Sen University, Zhuhai, 519082, People's Republic of China}
\affiliation{CSST Science Center for the Guangdong-Hong Kong-Macau Greater Bay Area, Zhuhai, 519082, People's Republic of China}
\affiliation{State Key Laboratory of Lunar and Planetary Sciences, Macau University of Science and Technology, Macau, People's Republic of China}
\email{}

\author[0000-0002-0447-7207]{Wei Zhong}
\affiliation{Institute of Science and Technology for Deep Space Exploration, Nanjing University, Suzhou, 215163, People's Republic of China}
\email{}

\author[0000-0002-9702-4441]{Wei Wang}
 \affiliation{CAS Key Laboratory of Optical Astronomy, National Astronomical Observatories, Chinese Academy of Sciences, Datun Road A20, Beijing 100101, China}
 \email{}

\author[0000-0002-8869-6510]{Jianheng Guo}
\affiliation{Yunnan Observatories,
Chinese Academy of Sciences, P.O. Box 110, Kunming 650011, China}
\affiliation{University of CAS, Beijing, China}
\affiliation{Key Laboratory for the Structure and Evolution of Celestial Objects, CAS, Kunming 650011, China.}
\email{}

\author[0000-0003-2278-6932]{Xianyu Tan}
\affiliation{Tsung-Dao Lee Institute \& School of Physics and Astronomy, Shanghai Jiao Tong University, Shanghai 201210, China}
\email{}

 \author[0000-0002-0378-2023]{Bo Ma}
 \affiliation{School of Physics and Astronomy, Sun Yat-Sen University, Zhuhai, 519082, People's Republic of China}
 \affiliation{CSST Science Center for the Guangdong-Hong Kong-Macau Greater Bay Area, Zhuhai, 519082, People's Republic of China}
 \email{}

\author{Ruyi Wei}
 \affiliation{Electric Information School, Wuhan University, Wuhan, 430072, People's Republic of China}
 \email{}

\author[0000-0003-0454-7890]{Cong Yu}
\affiliation{School of Physics and Astronomy, Sun Yat-Sen University, Zhuhai, 519082, People's Republic of China}
\affiliation{CSST Science Center for the Guangdong-Hong Kong-Macau Greater Bay Area, Zhuhai, 519082, People's Republic of China}
\affiliation{State Key Laboratory of Lunar and Planetary Sciences, Macau University of Science and Technology, Macau, People's Republic of China}
\affiliation{International Centre of Supernovae, Yunnan Key Laboratory, Kunming 650216, People's Republic of China}
\email{}

\begin{abstract}

Vertical mixing disrupts the thermochemical equilibrium and introduces additional heat flux that alters exoplanetary atmospheric temperatures.
We investigate how this mixing-induced heat flux affects atmospheric chemistry. 
Temperature increase in the lower atmosphere by the mixing-induced heat flux alters species abundances there and modifies those in the upper atmosphere through vertical transport.
In the lower atmosphere, most species follow thermodynamic equilibrium with temperature changes. 
In the upper layers,species mixing ratios depend on the positions of quenching levels relative to the regions exhibiting significant mixing-induced temperature variations. 
When the quenching level resides within such region (e.g. CO, $\rm CH_4$, and $\rm H_2O$ with strong mixing), the mixing ratios in the upper atmosphere are modified due to changes in the quenched ratios affected by the temperature variation in the lower atmosphere.
This alters the mixing ratio of other species (e.g. NO and $\rm CO_2$) through the chemical reaction network, whose quenching occurs in the region without much temperature change. 
The mixing ratios of $\rm CH_4$, $\rm H_2O$, and $\rm NH_3$ decrease in the lower atmosphere with increasing mixing heat flux, similarly reducing these ratios in the upper atmosphere.
Conversely, the mixing ratios of CO, $\rm CO_2$, and NO rise in the lower atmosphere, with CO and $\rm CO_2$ also increasing in the upper levels, although NO decreases.
Weaker host star irradiation lowers the overall temperature of the planet, allowing a smaller mixing to have a similar effect.
We conclude that understanding the vertical mixing heat flux is essential for accurate atmospheric chemistry modeling and retrieval.
\end{abstract}



\section{Introduction} 

Since the initial discovery of the exoplanet 51 Pegasi b \citep{1995Mayor}, more than 5,700 planets have been identified. 
Advances in observational technology have enhanced interest in exoplanetary atmospheres, an important aspect of planets.
Understanding atmospheric chemistry is crucial for comprehending the physical and chemical processes that occur in these atmospheres. 
Atmospheric properties are vital for evaluating the habitability of a planet (e.g., \citealt{1993Kasting,2016Shields}). 
Analyzing the abundance of elements such as carbon, nitrogen, oxygen, and sulfur through observational data offers critical insights into the formation and evolutionary history of planets (e.g., \citealt{2011Oberg,2021Turrini}. 
Advanced observational instruments such as the James Webb Space Telescope (JWST) have produced a wealth of precise data, substantially advancing our understanding of exoplanetary atmospheres (e.g., \citealt{2023Carter,2023Miles,2024Xue,2024Thao}).

Interpreting atmospheric observational data necessitates a comprehensive understanding of the diverse physical and chemical processes occurring within planetary atmospheres (e.g., \citealt{2021Tsai}).
 Vertical mixing is an essential dynamic process that enables the vertical transport of materials, thereby influencing chemical equilibria and modifying the mixing ratios of atmospheric species (e.g., \citealt{2001Ackerman, 2007Hubeny, 2018Gao}).
Observations of spectral characteristics can reveal this disequilibrium chemistry. Recently, evidence for vertical mixing has been identified.   
For example, \cite{2024Sing} reported the characteristics of vertical mixing when analyzing the JWST-NIRSpec transmission spectrum of WASP-107b.

The impact of vertical mixing on atmospheric chemistry depends on the planetary characteristics, particularly the atmospheric thermal profile and the intensity of vertical mixing (e.g., \citealt{2014Zahnle,2022Mukherjee,2024Mukherjee}).
At elevated temperatures, the chemical reaction timescale is shorter than the timescale of kinetic mixing, resulting in an atmospheric composition predominantly governed by thermal chemical equilibrium.
Conversely, under cooler conditions, the chemical timescale exceeds the kinetic mixing timescale, allowing vertical mixing to substantially affect the concentration of atmospheric species.
The quenching level is defined as the pressure where chemical and mixing timescales match. Above this quenching level, the abundance of species remains consistent with the abundance achieved at the quenching level (e.g., \citealt{1977Prinn,2018Tsai}).

Vertical mixing also contributes to the energy transport within the atmosphere and shapes the temperature profile. It generates a vertical downward heat flux in the radiative region (e.g., \citealt{Youdin2010, 2018Leconte, 2024Zhong}). 
This downward heat flux results in an increase in temperature and enlarges the apparent radius of the planet (e.g., \citealt{2017Tremblin,2019Sainsbury,2021Fortney}). 
In the radiative layer, vertical mixing is induced by atmospheric circulation \citep{1984maph...Holton} and breaking gravity waves \citep{1987JGR....Strobel}. 
The mixing heat flux modifies the atmospheric thermal profile, pushing the radiative-convective boundary to higher pressure levels, thus compressing the convective layer deeper within the atmosphere \citep{Youdin2010}. 
 \cite{2024Zhong,2025Zhang,2025Zhong_3}, hereafter denoted as \citetalias{2024Zhong}, \citetalias{2025Zhang}, and \citetalias{2025Zhong_3}, incorporate this mixing heat flux into the calculation of the radiative transfer equation, illustrating its intricate impact on the thermal structure of planetary atmospheres.
 
Previous studies have investigated how vertical material transport influences chemical disequilibrium in exoplanet atmospheres under varying planetary conditions. (e.g., \citealt {2020Fortney,2024Mukherjee}).
Variations in the atmospheric temperature profile impact the interaction between atmospheric chemistry and mixing transport.
To comprehensively investigate the effects of vertical mixing on atmospheric chemistry, it is essential to simultaneously consider both the material transport and the energy redistribution driven by vertical mixing.
The combination of material transport adds complexity to the role of vertical mixing in atmospheric chemistry. 
This paper incorporates mixing heat flux into atmospheric chemistry calculations to evaluate its effects.
We employ the VULCAN code \citep{2017Tsai, 2021Tsai} to model disequilibrium chemistry that incorporates material transport resulting from vertical mixing. Additionally, we calculate the atmospheric temperature profiles as described in \citetalias{2024Zhong} $\sim$ \citetalias{2025Zhong_3}, which include the mixing heat flux, and use these profiles as input for VULCAN.

The framework of this work is structured as follows:
\S \ref{method} outlines the models and code employed in this study.
\S \ref{result} details the changes in species mixing ratios that result from incorporating mixing heat flux.
Finally, \S \ref{conclusion} offers a comprehensive summary and discussion.

\section{Method}\label{method}
This section outlines the methodology for simulating the disequilibrium chemistry induced by vertical mixing and describes the approach to determine atmospheric temperature with mixing heat flux.
In \S \ref{vulcan}, we review the mass continuity equation that governs the behavior of particles within the atmosphere and introduce the VULCAN code applied in our calculations.
To incorporate the heat flux arising from vertical mixing, we adapt the concept of radiative equilibrium (RE) in atmospheric temperature calculation to what we define as radiative-mixing equilibrium (RME), as detailed in \S \ref{RME}.

\subsection{Disequilibrium chemistry}\label{vulcan}
Exoplanetary atmospheres are inherently dynamic and are influenced by various physical processes, including dynamical mixing, sedimentation, and photochemistry. 
These processes drive atmospheric chemistry away from chemical equilibrium.
Chemical kinetic models provide a framework that connects these planetary processes to chemical reactions. 
These models compute the temporal evolution of species number density to determine either the ultimate stable species distribution or its distribution at a specific moment \citep{2011Moses,2017Tsai,2021Kawashima}.
In one-dimensional systems, the mass continuity equation for a given species i is expressed as follows:

\begin{equation}\label{eq. continuity}
    \frac{\partial n_i}{\partial t}=\mathcal{P}_i-\mathcal{L}_i-\frac{\partial \phi_i}{\partial z},
\end{equation}

where $n_i$ represents the number density of the $i$th species, $t$ signifies time, and $z$ denotes the spatial coordinate in the vertical direction.
The symbols $\mathcal{P}_i$ and $\mathcal{L}_i$ denote the production and loss rates of the $i$th species within the chemical network, respectively.
When focusing exclusively on vertical mixing and ignoring other disequilibrium effects, such as photochemistry, the transport flux $\phi_i$ can be formulated as follows:
\begin{equation}
    \phi_i=-K_{\rm zz}n_{\rm tot}\frac{\partial \mathcal{X}_i}{\partial z},
\end{equation}
where $K_{\rm zz}$ is the so-called eddy diffusion coefficient.

We employ the VULCAN code \citep{2017Tsai, 2021Tsai} to track disequilibrium chemistry and calculate the mixing ratios of various atmospheric species across a range of eddy diffusion coefficients.
VULCAN has been extensively employed to model atmospheric chemistry across different planets \citep{2021Tsai, 2025Leung}
This code implements the Rosenbrock method \citep{Rosenbrock1963} to track the temporal evolution of species number densities and accounts for various disequilibrium processes, including photochemistry, condensation, particle settling, advection, eddy diffusion, and molecular diffusion. 
Given our focus on vertical mixing, we deactivate other disequilibrium processes and solve the mass continuity equation \eqref{eq. continuity} to obtain the number density ($\text{cm}^{-3}$) of atmospheric species.
For all scenarios, we applied the default ``NCHO thermo network".
The temperature profile, a crucial input for VULCAN, will be determined through a separate calculation.

\subsection{Radiative-mixing equilibrium}\label{RME}
This section outlines the methodology employed to calculate the atmospheric temperature profile.
Vertical mixing facilitates the transport of materials and generates heat flux.
In the convectively stable layer, vertical mixing contributes to a downward heat flux, which subsequently intensifies the greenhouse effect in high-pressure regions(\citealt{Youdin2010, 2018Leconte},  \citetalias{2024Zhong} $\sim$ \citetalias{2025Zhong_3}).
Given that atmospheric chemistry is temperature-dependent, it can be anticipated that the heat flux resulting from vertical mixing will modify the atmospheric chemical composition.
The heat flux induced by vertical mixing can be expressed using the following formula \citep{Youdin2010}: 

\begin{equation}\label{eq.Feddy}
    F_{\rm eddy}=-K_{\rm zz}\rho g\left(1-\frac{\nabla}{\nabla_{\rm ad}}\right) \ ,
\end{equation}
where $\rho$ represents the atmospheric density and $g$ denotes the gravitational acceleration.
$\nabla$ signifies the logarithmic temperature gradient, defined as follows:
\begin{equation}\label{nabla}
    \nabla= \frac{d\ln T }{d\ln P}=\frac{P}{T}\frac{dT}{dP} \ .
\end{equation}
 The adiabatic gradient for an ideal diatomic gas is given by $\nabla_{\rm ad} = 2/7$. $T$ and $P$ represent the temperature and pressure, respectively. 
 The intensity of the mixing heat flux is also quantified by the eddy diffusion coefficients $K_{\rm zz}$. 
When calculating the atmospheric temperature, the induced heat flux requires evolving from radiative equilibrium to radiative-mixing equilibrium, which is defined as follows (\citetalias{2024Zhong} $\sim$ \citetalias{2025Zhong_3}):
 \begin{equation}\label{nbalance1}
     \int^\infty_{\rm 0}\kappa_{\rm \nu}(J_{\rm \nu}-B_{\rm \nu})d\nu+\frac{ g}{4 \pi}\frac{d F_{\rm eddy}}{dP}=0 \ ,
 \end{equation}
$\kappa_{\rm \nu}$ represents the opacity and $B_{\rm \nu}$ denotes the Planck function. The first moment of the radiation intensity, $J_{\rm \nu}$, is given by:
\begin{equation}
    J_{\nu} =\frac{1}{2}\int^{1}_{\rm -1}I_{\rm \nu}(\mu)d\mu \ .
\end{equation} 
$I_{\nu}$ is the wavelength-dependent intensity at frequency ${\nu}$. In a steady-state, horizontally homogeneous atmosphere, $I_{\nu}$ satisfies the radiative transfer equation which is given by \citep{1960Chandrasekhar,2017Hengbook}: 
\begin{equation}
    \mu\frac{\partial I_{\rm \nu}}{\partial \tau_{\rm \nu}}=I_{\rm \nu}-S_{\rm \nu} \ .
\end{equation}
Here, $\mu \equiv \cos \theta$, where $\theta$  is the angle with respect to the vertical direction, and $\tau_{\nu}$ represents the optical depth. 
The source function $S_{\nu}$ is defined as the ratio of total emissivity to total opacity.
 For an in-depth description, refer to (\citetalias{2024Zhong} $\sim$ \citetalias{2025Zhong_3}). 
We obtain the atmospheric temperature profiles corresponding to various eddy diffusion coefficients via iterative computations. These profiles are subsequently utilized as input for the VULCAN code.

 When solving for atmospheric temperature, we employ the semi-grey approximation \citep{2010Guillot, Heng2014}, which divides opacity into two components: the visible band (denoted as ``v") for incoming irradiation and the thermal band (denoted as ``th") for outgoing emission. 
In the fiducial model, the gravitational acceleration, visible opacity, thermal scattering coefficient, and thermal opacity are defined as: $g = 10^3 \ \text{cm} \ \text{s}^{-2}$, $\kappa_{\rm v}=\sigma_{\rm th}=\kappa_{\rm th}=0.005\ \text{cm}^2\ \text{g}^{-1}$. 
The irradiation temperature is specified at $T_{\rm irr} = 1000 \ \text{K}$, corresponding to an equilibrium temperature of about $700 \ \text{K}$, which is similar to that of a warm Neptune, such as TOI-2407b \citep{2025Jan}. The internal temperature is set at $T_{\rm int} = 50 \ \text{K}$. In computational chemistry, only the elements C,H,O,N are considered, using elemental abundances from the protosolar composition as reported by \cite{2009Lodders}. These parameters generate an atmospheric temperature not exceeding 3000K, which is approximately within the applicable temperature range for the chemical reaction rates used in VULCAN.
Additionally, the cosine of the irradiation angle $\mu_{\rm *}$ is set to -1 to represent the substellar point.
The Eddington factors are consistent with those specified in \citetalias{2025Zhang}.

\section{Result}\label{result}
\begin{figure}
\includegraphics[width=\columnwidth]{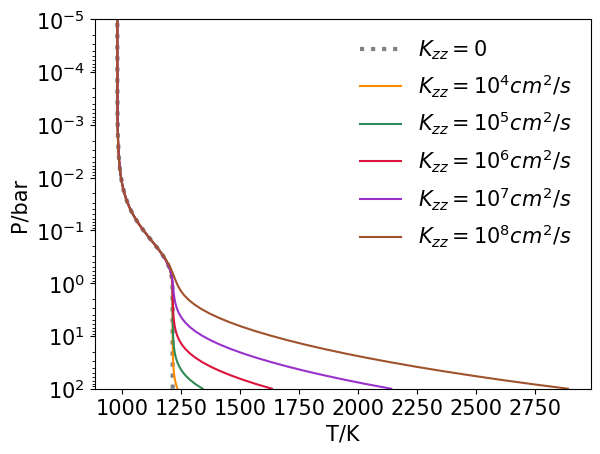}
    \caption{The temperature profiles for different values of $K_{\rm zz}$ under the fiducial model outlined in \S \ref{RME}. 
    }
    \label{figT0.005}
\end{figure}

The introduction of mixing heat flux results in the accumulation of energy in the lower atmosphere, causing a rise in temperatures within high-pressure areas, as illustrated in Figure \ref{figT0.005}. 
 This temperature shift affects the reaction rates of atmospheric constituents, thereby altering the proportions of various species at thermochemical equilibrium. 
Consequently, the mixing heat flux can transform the species mixing ratios, especially within the lower layers, where temperature experiences a notable increase due to this flux.
The influence of mixing heat flux on the upper atmospheric temperature is relatively minimal unless $K_{\rm zz}$ is substantial \citep{2024Zhong}. 
Nevertheless, variations in the number density of species within the lower and middle atmospheric layers, initiated by the mixing heat flux, can further impact the species mixing ratios in the upper atmosphere through vertical mixing-induced material transport.

Figure \ref{figT0.005} depicts the temperature profiles associated with varying eddy diffusion coefficients of the fiducial model. 
As the value of $K_{\rm zz}$ increases, the heating effect in the lower atmosphere becomes more pronounced, with significant temperature increases extending to higher altitudes. The pressure at which a noticeable change in temperature occurs decreases from approximately 10 bar at $K_{\rm zz}=10^{4} \ \text{cm}^{2}\ \text{s}^{-1}$ to about 1 bar at $K_{\rm zz}=10^{8} \ \text{cm}^{2}\ \text{s}^{-1}$.
We define the region below the pressure where the temperature begins to exhibit noticeable changes as the eddy flux heating region.

\subsection{The Effect of the Mixing heat flux}\label{res.eddy}
In the hot and dense regions of the atmosphere, chemical reactions occur on shorter timescales compared to vertical mixing, resulting in thermochemical equilibrium primarily determining the chemical distribution. 
The impact of mixing heat flux on the species mixing ratio in this region is mainly governed by how the chemical equilibrium abundances of species respond to temperature variations.
As pressure decreases, the timescale for chemical reactions gradually surpasses that of dynamic processes.
When focusing exclusively on vertical mixing and chemical equilibrium, vertical transport predominantly governs the species distribution in these areas.
Species become quenched and their mixing ratios are effectively frozen at values near the quenching levels.
When the quenching level is situated within the eddy flux heating region, modifications in the mixing ratio due to mixing heat flux in the lower atmosphere directly impact the mixing ratio above the quenching level.
In contrast, if the quenching level is positioned above the eddy flux heating region, the effect of mixing heat flux on species mixing ratios in the lower atmosphere does not directly affect the corresponding species mixing ratio in the upper atmosphere.
Instead, the mixing ratio of species in the upper atmosphere is influenced by other species whose quenching levels are within the heating region, mediated through the chemical network.
In the context of the fiducial model described in \S \ref{RME}, we will analyze several molecules as examples and discuss them according to the classification of the influence of mixing heat flux in the following section.

\subsubsection{$\rm CH_4$ and $\rm H_2O$}
\begin{figure*}[ht!]
\includegraphics[width=2\columnwidth]{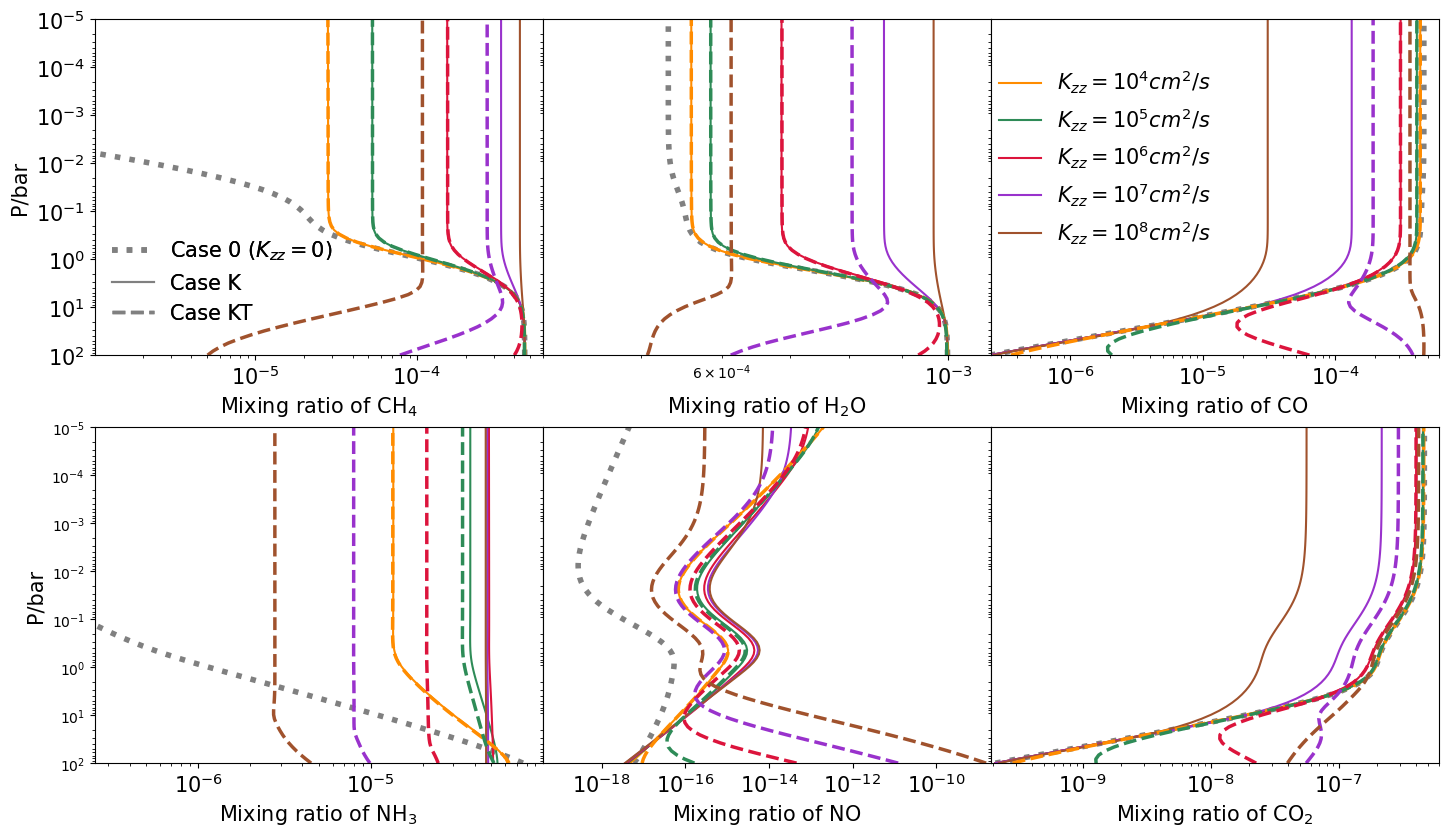}
    \caption{The mixing ratios of $\rm CH_4$, $\rm H_2O$, $\rm CO$, $\rm NH_3$, $\rm NO$ and $\rm CO_2$ across varying intensities of vertical mixing. The dotted line (Case 0) corresponds to the case where there is no vertical mixing in the atmosphere. The solid lines (Case K) represent the scenarios where only the effect of material transport due to vertical mixing is considered, without accounting for the mixing heat flux. The dashed lines (Case KT) consider both the material mixing and the heat flux brought about by vertical mixing. The different colors correspond to different values of $K_{\rm zz}$, with the temperature profile displayed in Figure \ref{figT0.005}.}
    \label{fig1000K}
\end{figure*}

This section examines the impact of mixing heat flux on methane ($\rm CH_4$) and water ($\rm H_2O$). 
Figure \ref{fig1000K} (a) illustrates the mixing ratio of $\rm CH_4$ across three distinct cases 
In Case 0, the vertical eddy diffusion coefficient is set to zero, implying that there is no vertical mixing in the calculations of both temperature and chemical composition
Case K incorporates the mass transport of vertical mixing while excluding its energy transport.
Case KT accounts for both heat flux and mass transport resulting from mixing.

Within the temperature range depicted in Figure \ref{figT0.005}, the concentration of $\rm CH_4$ in the lower atmosphere diminishes with increasing temperature at constant pressure. 
Consequently, its mixing ratio decreases as the value of $K_{\rm zz}$ rises, as shown in Figure \ref{fig1000K} (a).
In the fiducial model, the thermochemical equilibrium mixing ratio of methane increases with increasing pressure until it reaches a near-constant value at approximately 10 bar, as indicated by the dotted line.
 Comparing Case 0 and Case K reveals that vertical mixing-induced material transport raises the $\rm CH_4$ mixing ratio in the upper atmosphere.
 This process facilitates the upward transport of molecules from the lower atmosphere,  where $\rm CH_4$ has a high mixing ratio.
As the $K_{\rm zz}$ increases, $\rm CH_4$ becomes quenched at lower altitudes, which results in its mixing ratio in higher atmospheric layers incrementally increasing to resemble that found in the lower atmosphere.
 Enhanced vertical mixing promotes a more homogeneous distribution of methane across the atmospheric strata. 

The comparison between Case K and Case KT highlights the influence of mixing heat flux on the methane distribution.
When $K_{\rm zz}$ is small, the mixing heat flux exerts a weak heating effect, having a limited influence on the mixing ratio of $\rm CH_4$ in the lower atmosphere.  In these scenarios, the $\rm CH_4$ distribution remains relatively unchanged, as shown by the yellow and green lines in Figure \ref{fig1000K} (a).
As $K_{\rm zz}$ increases, notable transformations manifest. 
The temperature increase driven by the mixing heat flux contributes to a gradual reduction in methane concentration at the same pressure in the lower atmosphere.
This concept is consistent with scenarios depicted in Figures 5 and 6 of \cite{2017Tsai} and Figure 3 of \cite{2024Mukherjee}. 
Specifically, in high $K_{\rm zz}$ conditions, the methane mixing ratio is shown to be lower than in low $K_{\rm zz}$ conditions when the pressure is around 10 bar.
Simultaneously, the eddy flux heating region expands into the lower-pressure region with the increase of $K_{\rm zz}$, while chemical quenching occurs at deeper layers due to enhanced mixing. 
Therefore, the quenching level gradually shifts into the eddy flux heating region, amplifying the impact of mixing heat flux on species mixing ratios in the upper atmosphere.
Figure \ref{figT0.005} demonstrates that the temperature in the upper atmosphere remains predominantly unaffected by the eddy flux.
However, the diminished methane in the lower layers triggers a subsequent decline in methane concentrations in the upper layers through material transport, eventually decreasing the methane mixing ratio in the upper atmosphere as $K_{\rm zz}$ increases.
When $K_{\rm zz}=10^8\, \text{cm}^2\, \text{s}^{-1}$,  the $\text{CH}_4$ mixing ratio in the upper atmosphere is less than at $K_{\rm zz}=10^7\, \text{cm}^2\, \text{s}^{-1}$, with the peak mixing ratio observed near $K_{\rm zz}=10^7\, \text{cm}^2\, \text{s}^{-1}$. 
This finding suggests that the methane mixing ratio in the upper atmosphere does not rise continuously as $K_{\rm zz}$ increases, but instead decreases after surpassing a certain point.

Figure \ref{fig1000K} (b) demonstrates that $\rm H_2O$ exhibits a pattern analogous to $\rm CH_4$. The heat flux resulting from mixing in the lower atmosphere leads to a decline in the water mixing ratio.
Furthermore, this reduction also decreases the water mixing ratio in the upper layer when the parameter $K_{\rm zz}$ is sufficiently large. 
The maximum mixing ratio of $\rm H_2O$ in the upper atmosphere also occurs near $K_{\rm zz}=10^7 \, \text{cm}^2 \, \text{s}^{-1}$. 
In comparison to methane, water exhibits a weaker response to temperature changes and a smaller change in the mixing ratio caused by the mixing heat flux.
 Specifically, at $K_{\rm zz}=10^8\ \text{cm}^2\ \text{s}^{-1}$, the eddy flux diminishes the upper layer water mixing ratio to 60\% of what it would be with only mixing mass transport. For methane, this reduction is about 30\%.
 
\subsubsection{$\rm CO$ and $\rm CO_2$}\label{CO12}
This section discusses the impact of mixing heat flux on the mixing ratios of carbon monoxide ($\rm CO$) and carbon dioxide ($\rm CO_2$). 
In contrast to methane and water, CO exhibits a fundamentally different behavior. 
When focusing solely on the matter transport driven by vertical mixing, a decrease in the $\rm CO$ mixing ratio is observed in the upper atmosphere, as demonstrated by the solid lines in Figure \ref{fig1000K} (c).
In the lower atmosphere, as the atmospheric temperature rises caused by the mixing heat flux, the mixing ratio of $\rm CO$ at a specific pressure level increases. 
This aligns with the phenomena depicted in Figures 5 and 6 of \cite{2017Tsai} within the same pressure range.

Incorporating mixing heat flux into the analysis reveals that the associated temperature increase in the lower atmosphere results in heightened mixing ratios of CO in the upper atmosphere.
This is illustrated by comparing the solid and dashed lines in Figure \ref{fig1000K} (c).
For substantial $K_{\rm zz}$ values, such as $K_{\rm zz}=10^8\ \text{cm}^2\ \text{s}^{-1}$ and $K_{\rm zz}=10^7\ \text{cm}^2\ \text{s}^{-1}$, the quenching level of CO is located within the heating region of the eddy flux. 
In these cases, an increase in the mixing ratio within the lower atmosphere leads to a corresponding rise in the CO mixing ratio in the upper atmosphere. 
Notably, when $K_{\rm zz}=10^8\ \text{cm}^2\ \text{s}^{-1}$, the mixing heat flux enhances the CO mixing ratio in the upper atmosphere by approximately tenfold, restoring it to levels comparable to scenarios where vertical mixing is disregarded (Case 0).
 As the value of $K_{\rm zz}$ exceeds a certain threshold, the concentration of CO in the upper atmosphere begins to increase rather than decrease with increasing $K_{\rm zz}$.
 Specifically, when $K_{\rm zz}$ is approximately $10^7\ \text{cm}^2\ \text{s}^{-1}$, the concentration of CO in the upper atmosphere reaches its lowest.

Not all species exhibit quenching levels within the eddy flux heating region. 
As illustrated in Figure \ref{fig1000K} (f), the quenching levels for $\rm CO_2$ at varying values of $K_{\rm zz}$ are consistently located above about 0.1 bar. 
This indicates that heating from mixing heat flux in the lower atmosphere does not directly affect the mixing ratio of $\rm CO_2$ in the upper atmosphere.
In the middle and upper atmosphere, where temperature variations caused by mixing heat flux are negligible, alterations in the $\rm CO_2$ mixing ratio are still observed when $K_{\rm zz}$ is large. 
This is caused by variations in the mixing ratios of other chemical species, particularly CO, at corresponding altitudes through intricate chemical networks. 
Driven by CO, the mixing ratio of $\rm CO_2$ shows similar trends when mixing heat flux is present. 
In the upper atmosphere, the mixing ratio of $\rm CO_2$ does not decrease monotonically with increasing $K_{\rm zz}$ but instead reaches a minimum around $K_{\rm zz}=10^7\ \text{cm}^2\ \text{s}^{-1}$.

The inherent complexity of atmospheric chemical reactions produces intricate outcomes due to changes in the mixing ratios of various species.
Different species influence one another through the complex chemical reaction network. Consequently, $\rm CO_2$ does not change completely in synchrony with CO.
 Furthermore, the rate coefficients of chemical reactions are not monotonic functions of temperature, leading to non-monotonic changes in the mixing ratios of species with temperature. 
Consequently, the outcomes may vary if the atmospheric equivalent temperature is altered.

\subsubsection{$\rm NH_3$ and $\rm NO$}

This section discusses the results concerning the nitrogen compounds ammonia ($\rm NH_3$) and nitrogen monoxide (NO). 
In the fiducial model, ammonia is the most abundant nitrogen-containing species aside from $\rm N_2$.
Figure \ref{fig1000K} (d) illustrates that the impact of mixing heat flux on the mixing ratio of $\rm NH_3$ in the upper atmosphere resembles that of $\rm CH_4$. 
The reduction of $\rm NH_3$ in the lower layers leads to a corresponding decrease in the mixing ratio of $\rm NH_3$ in the upper layers through material transport.
 Ammonia exhibits heightened sensitivity to material transport and temperature, as demonstrated by \cite{2024Mukherjee}, with its quenching level occurring at lower altitudes. 
 A lower value of $K_{\rm zz}$ can enable the material transport process to play a notable role within the eddy flux heating region, thereby altering the ammonia mixing ratio. 
With the same $K_{\rm zz}$, the decrease in ammonia in the upper atmosphere due to mixing heat flux is also more pronounced. 
Specifically, when $K_{\rm zz}=10^{8}\ \text{cm}^2\ \text{s}^{-1}$, the mixing ratio of ammonia drops to nearly 5\% of its value observed in the case without the mixing heat flux.

Figure \ref{fig1000K} (e) illustrates the impact of mixing heat flux on the mixing ratio of $\rm NO$. 
In the lower atmosphere, the comparison between the solid and dashed lines demonstrates that an increase in temperature driven by mixing heat flux results in a higher concentration of $\rm NO$.
However, the upper atmosphere exhibits a decline in the $\rm NO$ mixing ratio,  which differs from the pattern observed in the lower layers.
 The quenching level of NO is located far from the eddy flux heating region.
Similar to $\rm CO_2$, in the middle layers where temperature variations due to mixing heat flux are minimal and material transport is inadequate, the mixing ratio of NO is influenced by other chemical species through chemical reaction when the mixing heat flux is considered.
The mixing ratio of NO in the middle layer has shown a similar decreasing trend to that of ammonia when considering the mixing heat flux.
 This reduction also contributes to a difference in the concentration of NO when it is quenched in the upper atmosphere.

 \subsection{The Effect of the Irradiation Temperature}\label{s.tirr}

This section explores the impact of irradiation temperature on how mixing heat flux affects atmospheric chemistry.
 We respectively varied the irradiation temperature while maintaining other parameters consistent with the fiducial model. 
For each specific irradiation temperature, the mixing ratio profiles for various species across several mixing levels is calculated by a method similar to \S \ref{res.eddy}.
We display the mixing ratios of different species at pressures of 1 mbar and 10 bar, representing the upper and lower atmospheric layers, respectively. 

\begin{figure}[ht!]
\includegraphics[width=\columnwidth]{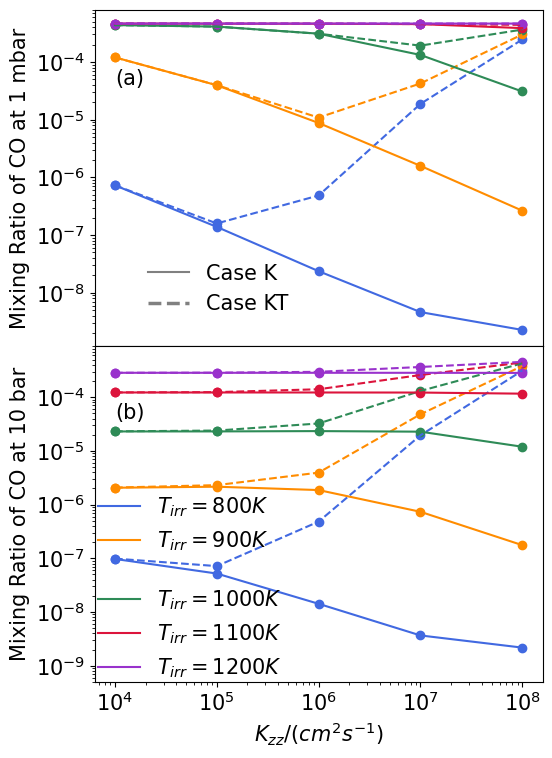}
    \caption{The first row shows the mixing ratios of CO at 1 mbar across varying intensities of vertical mixing, and the second row presents those at 10 bar. The solid lines (Case K) represent the scenarios where only the effect of material transport due to vertical mixing is considered, without accounting for the mixing heat flux. The dashed lines (Case KT) consider both the material mixing and the heat flux brought about by vertical mixing. The different colors correspond to different values of irradiation temperature $T_{\rm irr}$.}
    \label{fig.Tirr}
\end{figure}
Radiation from the host star plays a pivotal role in influencing the temperature of an exoplanet's atmosphere, with its strength often described by the irradiation temperature parameter $T_{\rm irr}$. To demonstrate the impact of irradiation temperature on our results, we utilize CO as an example. Figure \ref{fig.Tirr} presents the mixing ratios of $\rm CO$ at pressure levels of 1 mbar and 10 bar for a range of irradiation temperatures $T_{\rm irr}$. Changes in the intensity of radiation the planet receives lead to variations in its temperature, which in turn affect atmospheric chemistry, as depicted by the various colored lines in Figure \ref{fig.Tirr}.

Decreasing irradiation temperature results in lower atmospheric temperatures, lengthening the timescale of chemical reactions. 
 Consequently, the atmosphere is more susceptible to being driven out of thermochemical equilibrium by mixing transport.
 In the absence of mixing heat flux, as depicted by the solid line in Figure \ref{fig.Tirr} (b), for $T_{\rm irr}$ values of 800, 900, or 1000 K, the CO mixing ratio at 10 bar reduces due to enhanced mixing, despite the temperature remaining constant as $K_{\rm zz}$ increases.
 This drop in the CO mixing ratio is attributed to material transport. 
 As $T_{\rm irr}$ decreases, this drop becomes more significant.
 In contrast, for higher $T_{\rm irr}$, such as 1100 and 1200 K, the CO mixing ratio at 10 bar remains constant despite increasing $K_{\rm zz}$. 
 At higher temperatures, the shorter chemical reaction timescale necessitates a more pronounced mixing effect to alter the CO mixing ratio, beyond the depicted $K_{\rm zz}$ values.
 The CO mixing ratio at 1 mbar in the upper atmosphere also decreases in line with the reduction seen in the lower atmosphere, as indicated by the corresponding solid line in Figure \ref{fig.Tirr} (a).
 
The dashed lines illustrated in Figure \ref{fig.Tirr} indicates the results that incorporate the mixing heat flux. The comparison between the solid and dashed lines underscores the influence of the mixing heat flux. At a pressure of 10 bar, this flux causes a temperature rise, which in turn increases the CO mixing ratio, akin to the findings depicted in Figure \ref{fig1000K}. This effect is more noticeable on planets with less irradiation.
It is important to note that different species react distinctly to temperature fluctuations, so not all follow the same trend. 
As stated in \S \ref{CO12}, an elevated CO concentration in the lower atmosphere due to the mixing heat flux can escalate the mixing ratio in the upper atmosphere via material transport when $K_{\rm zz}$ is sufficiently large. 
With a reduction in irradiation temperature, the CO mixing ratio in the upper atmosphere can notably rise due to the mixing heat flux even at relatively low $K_{\rm zz}$ values.
Concurrently, the CO mixing ratio in the upper atmosphere hits a minimum at a particular $K_{\rm zz}$ value under Case kt, and this minimum appears at a lower $K_{\rm zz}$ when the irradiation temperature is lower.
The reason is that as the atmospheric temperature decreases with reduced irradiation temperature, the timescale for chemical reactions becomes longer, allowing lighter mixing to have similar effects. 
Conversely, with increased irradiation temperature, the overall atmospheric temperature heightens, reducing the timescale for chemical reactions. 
In such a case, only more intense mixing allows species to stabilize at lower altitudes. 
Changes in the mixing ratio of the lower atmosphere can affect the upper atmosphere's mixing ratio only under conditions of more vigorous mixing.

The dashed lines in Figure \ref{fig.Tirr} represents the results incorporating the mixing heat flux.
The comparison between the solid and dashed lines highlights the impact of the mixing heat flux.
At a pressure of 10 bar, this flux causes a temperature rise, which in turn increases the CO mixing ratio, similar to what is shown in Figure \ref{fig1000K}.
This effect is more noticeable on planets with less irradiation.   
It is important to note that different species react distinctly to temperature fluctuations,  so not all species follow the same trend.
As mentioned in \S \ref{CO12}, the increase in CO concentration in the lower atmosphere due to the mixing heat flux can escalate the mixing ratio in the upper atmosphere via material transport when $K_{\rm zz}$ is sufficiently large.
With decreasing irradiation temperature, the CO mixing ratio in the upper atmosphere can notably rise due to the mixing heat flux even at relatively low $K_{\rm zz}$ values.
Simultaneously, the CO mixing ratio in the upper atmosphere reaches its minimum at a specific $K_{\rm zz}$ value under Case kt, and this minimum occurs at a smaller $K_{\rm zz}$ when the irradiation temperature is smaller. 
The reason is that as the atmospheric temperature decreases with reduced irradiation temperature, the timescale for chemical reactions becomes longer, allowing lighter mixing to have similar effects.
 Conversely, with increased irradiation temperature, the overall atmospheric temperature heightens, reducing the timescale for chemical reactions. 
In such a case, only more intense mixing allows species to quench at lower altitudes. 
Changes in the mixing ratio of the lower atmosphere can only affect the mixing ratio in the upper atmosphere under conditions of stronger mixing.


\section{Discussion}
When addressing vertical mixing, we assume that the mixing occurs at a scale much smaller than the atmospheric pressure scale height. This assumption allows us to model the atmospheric dynamics as a diffusion process, employing the eddy diffusion coefficient.
However, the reality is far more intricate. As noted in \cite{2017Tsai}, three-dimensional models of hot Jupiter atmosphere demonstrate that the atmospheric circulation can extend over several orders of magnitude in vertical pressure (e.g., \citealt{2013Parmentier,Mendon2016}). 
This complexity challenges the reliability of the eddy diffusion approximation. Nevertheless, such approximation is still widely utilized in atmospheric chemistry (e.g., \citealt{1981Allen,2011Moses,2016Rimmer,2017Tsai}). 
The coefficient $K_{\rm zz}$ represents the cumulative capabilities of large- and small-scale mixing processes in material transport.
However, the formula for mixing heat flux, as outlined in Equation \eqref{eq.Feddy}, is limited to small-scale turbulent mixing that adheres to entropy conservation \citep{Youdin2010}. 
This restriction implies a distinction between the coefficient $K_{\rm zz,M}$, which measures material transport, and the coefficient $K_{\rm zz,E}$, which measures heat flux.
\cite{Youdin2010} found that the upper limit for the turbulent diffusion coefficient under typical hot Jupiter conditions is several orders of magnitude lower than the values $K_{\rm zz}$ related to chemical species mixing reported in other studies (e.g., \citealt{2009Spiegel,2009Zahnle}).
Different patterns of vertical atmospheric motion will alter the relationship between $K_{\rm zz,M}$ and $K_{\rm zz,E}$. 
The divergence between these two coefficients introduces an additional free parameter into the exoplanetary atmosphere, which is likely variable with altitude.

Several studies have estimated the value of $K_{\rm zz}$ based on observational data from the atmospheres of brown dwarfs (e.g., \citealt{2020Miles,2024Kothari}) and exoplanets (e.g., \citealt{2021Kawashima,2024Welbanks,2024Sing}). 
Various modeling efforts have sought to determine the profile of $K_{\rm zz}$ in the atmosphere (e.g., \citealt{2019Komacek,2022Tan,2023Arfaux}).
These studies indicate a diverse range of $K_{\rm zz}$ values, which are expected to vary under different planetary atmospheric conditions. 
Examining the value of 
$K_{\rm zz,M}$ and $K_{\rm zz,E}$ presents a complex challenge and will be addressed in detail in future work.
Consequently, we do not engage in a detailed discussion regarding the specific values of $K_{\rm zz}$, as this does not influence our qualitative assessment of the impact of mixing heat flux on atmospheric chemistry. 
For simplicity, we assume that both $K_{\rm zz,M}$ and $K_{\rm zz,E}$ are equal and set them to a constant value.

This study illustrates the complex influence of mixing heat flux on atmospheric chemistry.
Different chemical species exhibit varying responses to the changes in atmospheric temperature, resulting in distinct effects of the mixing heat flux. 
The non-linear relationship between the mixing ratio of chemical species and temperature means that employing atmospheric parameters distinct from those described in \S \ref{RME} and \S \ref{s.tirr} could produce diverse outcomes. 
For instance, the mixing ratio of species in the lower atmosphere may not vary monotonically with increasing $K_{\rm zz}$ values.

 Exoplanets exhibit a wide range of elemental abundances, such as the super-stellar C/O ratio (e.g., \citealt{2025Evans-Soma}), which influence their atmospheric chemistry. 
We tested the impact of the change in the C/O ratio and found that the C/O ratio did not have a significant effect on the processes that the mixing heat flux affects the species mixing ratios of the lower layer, and then changes those of the upper layer.
Nevertheless, variations in the distribution of different chemical species resulting from the C/O ratio (e.g., \citealt{2017Tsai}) can introduce complexities in the details of how chemical components respond to mixing heat flux.
These nonlinearity further complicates the changes in the material mixing ratio in the upper atmosphere. 
Therefore, specific calculations must be conducted when studying the atmosphere of a particular planet.

The effect of mixing heat flux on the mixing ratio of chemical species in the upper atmosphere enables the determination of the presence of this flux through observational data. Variations in the concentrations of chemical species directly modify the atmospheric spectrum. Future research will investigate the impact of mixing heat flux on this spectrum.

\section{conclusion}\label{conclusion}
Vertical mixing is vital for particle transport in the atmosphere and introduces an accompanying heat flux. 
The heat flux increases the temperature of the lower atmosphere, subsequently altering atmospheric chemistry.
To explore this effect, VULCAN was employed to simulate atmospheric chemistry under conditions both with and without the mixing heat flux. 
Atmospheric temperature profiles were derived following the methodology outlined in Paper \cite{2025Zhang} and were subsequently integrated into VULCAN.
The mixing heat flux modifies the chemistry of the lower atmosphere by changing temperature, which in turn affects upper atmospheric chemistry through particle transport.
Under our parameter settings, the mixing heat flux has negligible effects on low-pressure atmospheric temperatures. 
 The temperature profile and the spatial relationship between the chemical quenching levels and the eddy flux heating region are pivotal in assessing the impact of the mixing heat flux on species mixing ratios in the upper atmosphere.

The specific conclusions can be summarized as follows:
\begin{enumerate}
      \item The heat flux induced by vertical mixing raises the temperature in the lower atmosphere.
      This temperature variation subsequently alters the thermochemical equilibrium and consequently modifies the mixing ratios of species within this region.
      \item In the fiducial model, the temperature at nearly 10 bar progressively increases from approximately 1200 K to around 2000 K as the value of $K_{\rm zz}$ rises.
      Correspondingly, the concentrations of $\rm CH_4$, $\rm H_2O$, and $\rm NH_3$  decrease at this pressure, while the CO and NO mixing ratios increase with the ascent in $K_{\rm zz}$. 
      In the upper atmosphere, when vertical mixing is sufficiently pronounced, the mixing ratios of $\rm CH_4$, $\rm H_2O$, $\rm NH_3$, and NO are reduced due to the mixing heat flux, with ammonia exhibiting the most significant decrease.
      Conversely, the mixing ratios of CO and $\rm CO_2$ increase as a consequence of the mixing heat flux.
      \item For large values of $K_{\rm zz}$, the quenching level is situated in the region that is significantly heated by the mixing heat flux, which we refer to as the eddy flux heating region. The mixing ratio at the quenching level is influenced by temperature changes. Above the quenching level, the species mixing ratio is frozen near the value at the quenching level.
      Although the mixing heat flux exerts minimal impact on the temperature of the upper atmosphere, the species mixing ratios in this region will also adjust in response to alterations in the mixing ratios at the quenching level.
      \item  When the quenching levels of certain species are located within the eddy flux heating region, variations in their mixing ratios in the upper atmosphere can influence the mixing ratios of other species through the chemical network, even though the quenching levels of those species are situated above the eddy flux heating region.
       \item When a planet experiences reduced irradiation from its host star, its overall atmospheric temperature decreases, extending the chemical reaction timescale. 
       This extended timescale allows the modified mixing ratios in the lower atmosphere, driven by heat flux, to affect the mixing ratio in the upper atmosphere, though with a lesser mixing intensity.
\end{enumerate}
In the discussion, we also highlighted the possibility of disparities between the material transfer capacity and energy transfer capacity during the mixing process. This would necessitate a detailed investigation into two separate eddy diffusion coefficients, $K_{\rm zz,M}$ for quantifying material transfer and $K_{\rm zz,E}$ for quantifying heat transport.

\begin{acknowledgments}
This work is supported by the National Natural Science Foundation of China (NSFC, No.12288102), the National SKA Program of China (grant No. 2022SKA0120101), the National Key R \& D Program of China (No. 2020YFC2201200), the science research grants from the China Manned Space Project (No. CMS-CSST-2021-B09, CMS-CSST-2021-B12, and CMS-CSST-2021-A10), International Centre of Supernovae, Yunnan Key Laboratory (No.202302AN360001), Guangdong Basic and Applied Basic Research Foundation (grant 2023A1515110805), and the grants from The Macau Science and Technology Development Fund, and opening fund of State Key Laboratory of Lunar and Planetary Sciences (Macau University of Science and Technology) (Macau FDCT Grant No. SKL-LPS(MUST)-2021-2023). C.Y. has been supported by the National Natural Science Foundation of China (grants 11521303, 11733010, 11873103, and 12373071).J.H. Guo has been supported by the National Natural Science Foundation of China (NSFC, No.12433009).
\end{acknowledgments}

\bibliography{sample7}{}
\bibliographystyle{aasjournalv7}



\end{document}